\documentclass[fleqn,a4paper,12pt]{article}
\usepackage{amssymb}
\usepackage{amsmath}
\usepackage[dvips]{graphicx}
\usepackage{placeins}
\usepackage{lscape}
\usepackage{a4}
\usepackage{epsfig}

\newcommand{\ri}{{\mathrm i}}
\newcommand{\p}{\partial}
\newcommand{\bea}{\begin{array}}
\newcommand{\eea}{\end{array}}

\catcode`@=11 \long
\def\@caption#1[#2]#3{\par\addcontentsline{\csname
ext@#1\endcsname}{#1} {\protect\numberline{\csname
the#1\endcsname}{\ignorespaces #2}} \begingroup \small
\@parboxrestore \@makecaption{\csname fnum@#1\endcsname}
{\ignorespaces #3}\par \endgroup} \catcode`@=12

\newcommand{\la}{\label}

\catcode`@=11 \long
\def\@caption#1[#2]#3{\par\addcontentsline{\csname
ext@#1\endcsname}{#1} {\protect\numberline{\csname
the#1\endcsname}{\ignorespaces #2}} \begingroup \small
\@parboxrestore \@makecaption{\csname fnum@#1\endcsname}
{\ignorespaces #3}\par \endgroup} \catcode`@=12

\begin{document}

\allowdisplaybreaks
 \begin{titlepage} \vskip 2cm

\begin{center} {\Large\bf Group classification of
Schr\"odinger equations with position dependent mass}
\vskip 3cm {\bf {A.
G. Nikitin }\footnote{E-mail: {\tt nikitin@imath.kiev.ua} }  \vskip 5pt {\sl Institute of Mathematics, National
Academy of Sciences of Ukraine,\\ 3 Tereshchenkivs'ka Street,
Kyiv-4, Ukraine, 01601\\}}

\vskip .2cm

{\bf T.M. Zasadko}\footnote{E-mail: {\tt tacita@ukr.net} }
\vskip 5pt
{\sl Taras Shevchenko National University of Ukraine,\\64 Volodymirska Street, Kyiv-4, Ukraine\\}
\end{center} \vskip .5cm \rm

\begin{abstract}
Maximal kinematical invariance groups of $2d$ Schr\"odinger equation with a position dependent mass and arbitrary potential are classified.
It is demonstrated that there exist seven classes of such equations possessing  non-equivalent continuous symmetry group. Three of these classes include arbitrary functions while the remaining ones are defined up to arbitrary parameters. In particular, for the case of  a constant mass the class missing in the Boyer classification (Boyer C P 1974  Helv. Phys. Acta{\bf 47}, 450) is indicated. A constructive test of (non)equivalence of a PDM system to a constant mass system is proposed.
\end{abstract}
\end{titlepage}
\section{Introduction\label{int}}
Group classification of differential equations consists in the
specification of their classes which  are characterized
by non-equivalent symmetry groups. It is an important
research field which has both fundamental and application values.

One of the most famous results in this field consists in the group
classification of Schr\"odinger equations  with arbitrary potentials,
which was carried out as  far back as in the seventieth of the
previous century in papers \cite{Hag}, \cite{Nied}, \cite{And},
\cite{Boy}. In particular, it was shown in  \cite{Hag} and
\cite{Nied} that in addition to the Galilei invariance, the free
Schr\"odinger equation (SE) admits conformal transformations of
independent variables provided the wave function is multiplied by a specific phase
factor dependent on these variables and transformation parameters. These results form group-theoretical grounds of quantum mechanics and give rise for many inspiring physical theories such
as the conformal quantum mechanics. The completed description of
possible symmetry groups of SE  and the
corresponding non-equivalent versions of these equations presented
in \cite{Hag}, \cite{Nied}, \cite{And}, \cite{Boy} is a
cornerstone of modern quantum physics.

More contemporary results concerning  the group classification of
the {\it nonlinear} Schr\"odinger and Ginsburg-Landau equations can be found in papers \cite{pop} and \cite{N1}, higher symmetries of non-stationary linear and non-linear SE were discussed in \cite{FN}, Lie symmetries of supersymmetric SE were studied in \cite{Bec1} and \cite{Bec2}. Finally, symmetries of the free SE in the non-commutative plane are presented in \cite{SGM}.

On the other hand, the group classification of SE with position dependent mass (PDM) continues to be an open problem, which is  a rather curious fact taking into account the fundamental role played by such equations in various branches of modern physics. There is a number of papers devoted to SE with various particular and
sometimes exotic symmetries, see, e.g., \cite{11}, \cite{rac},
\cite{Kal}, \cite{Robin}. But, at the best of our knowledge, even the standard symmetries of these
equations with respect to continuous transformation groups never
were investigated completely.

An attempt at a  systematic study of symmetries of
the PDM SE has been made in paper \cite{NZ}
where all such (stationary) 3$d$ equations with different symmetry
groups were classified. The number of non-equivalent PDM SE
appears to be rather extended and includes 16 classes. Seven of them
are defined up to arbitrary parameters while the remaining ones
include arbitrary functions. Rather surprisingly one of the
considered equations is invariant with respect to the 6-parameter
Lie group locally isomorphic to the Lorentz group.

In the present paper we make a group classification of time
dependent PDM SEs with 1+2 independent
variables, which will be referred as $2d$ SE. This is not only the first step to classification of equations with higher dimensions, but also an interesting self-consistent problem.  First, the planar systems are of particular interest in modern physics. Secondly, such equations naturally appear as reduced versions of multidimensional systems written in partially separated variables. One more origin of $2d$ SE with position dependent masses are quantized versions of classical Hamiltonian systems defined in Riemannian space in two dimensions, see, e.g., \cite{Kal} and references cited therein.

The classification problem in 1+2 dimensional space is nothing like as simple as it looked. In spite of the small number
of dependent and independent variables, there are complications  connected with  equivalence
 relations which are much more powerful than in $3d$ case. The considered systems include two arbitrary elements, i.e., mass and potential terms, which make the classification problem rather nontrivial.

Since the ordinary SE with a constant
mass appears in our analysis as a particular case , we revised also
the classification results presented in \cite{Boy}.
We did not plane to verify them. However, some systems appearing in our analysis are equivalent to systems with  constant masses, and willy-nilly we were supposed to recover some of such systems. To our astonishment, in this way we discover a system  missing in the list presented in \cite{Boy}, see Section \ref{neww} below.

\section{Time dependent PDM Schr\"odinger equations}
The subject of our analysis are PDM Schr\"odinger equations of the
following generic form
\begin{gather}\la{seq}L\psi\equiv\left(\ri\frac{\p}{\p
t}-H\right)\psi=0\end{gather} where $H$ is the PDM Hamiltonian:
\begin{gather}\la{H1}  H=H_k+ V\end{gather} where
\begin{gather}\la{N1}H_k=\frac14(m^\alpha p_a m^{\beta}p_am^\gamma+m^\gamma p_a m^{\beta}p_am^\alpha).\end{gather}
Here $p_a=-\ri\frac{\p}{\p x_a}$, $m=m({\bf x})$ and $V=V({\bf x})$
are the mass and potential depending on spatial variables ${\bf
x}=(x_1,x_2,...,x_n)$, and summation w.r.t. the repeating indices
$a$ is imposed over the values $a=1,2,...,n$. This convention will be used in all the following text.

In (\ref{H1}) we use the physically motivated representation
(\ref{N1}) for the kinetic energy term, where $\alpha, \beta$  and $\gamma$ are the
ambiguity parameters satisfying the condition $\alpha+\beta+\gamma=-1$ \cite{Roz}.

Let us note that it is not the most general form of $H_k$ compatible with the hermicity condition, see interesting discussion of this point in \cite{Tra}. On the other hand, there are several reasons to reduce $H_k$ to more simple form by considering only  particular values  of the ambiguity parameters. Mathematically, it is possible to represent the PDM Hamiltonian in
the following forms equivalent to (\ref{H1}):
\begin{gather}\la{H2}H= p_a fp_a+ \tilde V\end{gather}
 and
\begin{gather}\la{H3}H= \sqrt{f}p_a p_a\sqrt{f}+ \hat V\end{gather}
where
\begin{gather}\la{H2-1}f=\frac1{2m},\qquad\tilde
V=V+\alpha\gamma\frac{f_af_a}{f}+\frac{\alpha+\gamma}2f_{aa}\end{gather}
 and
\begin{gather}\la{H2-3}\hat V= V+\left(\alpha\gamma-\frac14\right)\frac{f_af_a}{f}
+\frac{1+\alpha+\gamma}2f_{aa}\end{gather}
correspondingly.
Here $f_a=\frac{\p f}{\p x_a}$ and $f_{aa}=\frac{\p f_a}{\p x_a}$.

Hamiltonians (\ref{H2}) and (\ref{H3}) are simple equal to
 (\ref{H1}) but written in more compact forms convenient
for investigation of their symmetries.

Just representation
(\ref{H2}) was first proposed in \cite{ben} as the simplest way to  to
ensure current conservation. Representation (\ref{H2}) was applied  in \cite{Zhu} for reformulating the connection rule problem on
the two sides of an heterojunction. The same representation was derived in \cite{Cav} as the nonrelativistic limit of the Dirac Hamiltonian with PDM, and in \cite{yung} via path integral evaluation.

There are also various  other  versions of Roos ordering (\ref{N1}) which are suitable for particular  physical models: $\alpha=\gamma$ \cite{mora}, $\beta=\alpha=0, \gamma=-1$ \cite{gor}, $\gamma=0, \alpha=\beta=-\frac12$ \cite{li}, $\alpha=\gamma=-\frac14, \beta=-\frac12$ \cite{mustafa}. All the corresponding Hamiltonians (\ref{H1}) can be reduced to form (\ref{H2}) or (\ref{H3}), and so are equivalent up to redefinitions of the potential given by formulae  (\ref{H2-1}) and (\ref{H2-3}). On the other hand, different orderings can request different ways of introducing interaction (e.g., with the external electromagnetic field) into the corresponding models, and so they are not equivalent physically.

Representation (\ref{H2}) was used in \cite{NZ} for classification of first order
integrals of motion for PDM Hamiltonians, while representation
(\ref{H3}) appears to be convenient for classification of second
order integrals of motion \cite{N2}. Notice that all representations
of Hamiltonians given above are formally self adjoint.

Finally, let us transform Hamiltonian (\ref{H3}) and solutions of
equation (\ref{seq}) in the following manner:
\begin{gather}\la{H4}\psi\to \tilde \psi=f^{-\frac12}\psi,\quad H\to\hat
H=f^{-\frac12}Hf^{\frac12}=fp_ap_a+\hat V.\end{gather}
Representation (\ref{H4}) was used in \cite{N2} to construct
exact solutions of the corresponding eigenvalue problems.

In the present paper we restrict ourselves to group classification
of equations (\ref{seq}) with two  spatial variables.  Moreower, we will study Hamiltonians in the form (\ref{H2}). The
classification results can be easily reformulated for systems with
Hamiltonians represented in (\ref{H1}), (\ref{H3}) and (\ref{H4}),
since the relations between these representations are rather
straightforward and given by formulae (\ref{H2-1}), (\ref{H2-3}) and
(\ref{H4}).

We will not consider systems with one spatial dimension, since using
the Liouville transformation they always can be  reduced to systems
with constant masses, see \cite{N2} for examples and references. The
first steps of our analysis presented in the following section are valid for systems of arbitrary dimension $n$.

\section{Determining equations for symmetries of systems (\ref{seq}) with any dimension}
Let us search for symmetries of equations (\ref{seq}) with respect
to continuous groups of transformations. It can be done using the
classical Lie algorithm whose
contemporary version can be found in monograph \cite{olver}. In
application to the linear Schr\"odinger equation this algorithm can
be reduced to searching for the first order differential operators
of the following form:
\begin{gather}\label{so}
    Q=\xi^0\partial _t+\xi^a\partial_a+\tilde\eta\equiv \xi^0\partial _t+
    \frac12\left(\xi^a\partial_a+\partial_a\xi^a\right)+\ri\eta,
\end{gather}
associated with the group generators. In (\ref{so}) $\tilde
\eta=\frac12\xi^a_a+\ri\eta,\ \ $ $\xi^0$, $\xi^a$ and $\eta$ are
functions of independent variables, whose explicit form can be found
from the following operator equation
\begin{gather}\la{ic}QL-LQ=\alpha L\end{gather}
where $\alpha$ is one more unknown function of $t$ and $\bf x$.

Evaluating commutator in (\ref{ic}) and equating coefficients for
the same differentials we come to the following system of the
determining equations for unknowns $\xi^0, \xi^a, \eta, f, \hat V $
and $\alpha$:
\begin{gather}\label{de1}
\dot\xi^0=-\alpha, \quad  \xi^0_a=0 ,\\{\cal F}^{ab}\equiv
\left(\xi^b_{a}+\xi^a_{b}\right)f-\delta^{ab}\left(\xi^if_i-\alpha
f\right)=0,\label{de2}\\\label{de3} -i\dot\xi^a+f\xi^a_{cc}-2\ri
f\eta_{a}+\xi^bf_{ab}-\xi^a_bf_b-f\xi^n_{na}=\alpha
f_a,\\\label{de4}
 \ri (f\eta_{a})_a+\xi^aV_{a}+\frac12\left(f\xi^a_{ab}\right)_b+
 \left(\frac{\ri}2\dot\xi^a_{a}-
 \dot\eta\right)=\alpha V
\end{gather}
where $\delta^{ab}$ is the Kronecker symbol, the dot and subindices denote
derivations with respect to time and spatial variables:
$\dot \xi^0=\frac{\p \xi^0}{\p t}, \ f_{a}=\frac{\p f}{\p x_a}$, etc.

Considering various differential and algebraic consequences of
(\ref{de2})--(\ref{de4}) it is possible to reduce this system  to
the following equivalent form (see Appendix 1):
\begin{gather}\la{de5} \xi^b_{a}+\xi^a_{b}-\frac{2}n\delta_{ab}\xi^i_i=0,
\\\label{de6} \xi^if_i-\alpha f=\frac2n f\xi^i_i,\\ \dot\xi^a+2\eta_a
f=0,\label{de7}\\
\label{de8}\xi^aV_a+\frac12\xi^b_{ba}f_a=\alpha
V+\dot\eta.\end{gather}

The system (\ref{de1}), (\ref{de5})-(\ref{de8}) is overdetermined
but rather complicated. It can hardly be solved for arbitrary $n$ if
at all. In the following sections we present solutions of this
system for $n=2$.

\section{Reduction to the case n=2 and equivalence relations}

The classification problem in 1+2 dimensional space is rather
specific. First, the number of dependent and independent variables
is relatively small, secondly, the equivalence relations appear to be
more  powerful than in the cases of more extended carrier spaces.

Equation (\ref{de5}) for $n=2$ is reduced to the Caushy-Riemann
conditions for functions $\xi^1=u$ and $\xi^2=v$:
\begin{gather}{u}_1=v_2,\quad u_2=-v_1.\la{CR}\end{gather}
The corresponding symmetries (\ref{so}) can be rewritten as follows:
\begin{gather}\la{so1}Q=\xi^0\p_0+u\p_1+v\p_2+u_1+i\eta.\end{gather}
The remaining equations (\ref{de6})--(\ref{de8}) take the following
form:
\begin{gather}2f\eta_1+\dot u=0,\la{de11}\\
2f\eta_2+\dot v=0,\la{de12}\\
uf_1+vf_2-(\alpha+2u_1)f=0,\la{de13}\\
uV_1+vV_2+u_{11}f_1+v_{22}f_2=\alpha V +\dot\eta.\la{de14}
\end{gather}

Just system (\ref{de1}), (\ref{CR}), (\ref{de11})--(\ref{de14}) together with definition (\ref{so1})  describes all possible generators of
continuous symmetry groups and the corresponding position dependent
masses and potentials present in equations (\ref{seq}), (\ref{H2}) for
$n=2$. We will classify equations (\ref{seq}), (\ref{H2}) with
different symmetries up to equivalence relations defined in the
following.

Let us note that Hamiltonian (\ref{H2}) is form invariant with
respect to the following transformation including simultaneous changes of dependent and independent variables:
\begin{gather}\la{Change1} x_1\to \tilde u(x_1,x_2),\qquad x_2\to
\tilde v(x_1,
x_2),\\\la{Change2} \psi\to \frac1{\sqrt{R}}\psi,\qquad H\to\frac1{\sqrt{R}}H\sqrt{R}\end{gather} where $\tilde u$ and $\tilde v$ are arbitrary functions satisfying the Caushy-Riemann condition (\ref{CR}), and $R=\sqrt{\tilde u_1^2+\tilde u_2^2}.$ Indeed, transformations (\ref{Change1}) and (\ref{Change2}) keep the generic form of the Hamiltonian but change functions $f$ and $V$ in (\ref{H2}).

In addition,
equation (\ref{seq}) admits the scaling and shifts of the time
variable:
\begin{gather}\la{change}t\to \nu t+\mu, \ \end{gather}
were $\nu\neq0$ and $\mu$ are arbitrary constants. Such
transformations can be compensated by multiplication of $f$ and $V$
by $\frac1\nu$.

It is possible to show that formulae (\ref{Change1}), (\ref{Change2}) and (\ref{change}) give  the most general continuous  transformations which keep the generic form (\ref{H2}) of the Hamiltonian up to change of functions $f$ and $V$.
The validity of this statement can be verified by a direct calculation.
Notice that just transformations of type (\ref{Change1}) correspond to generic infinitesimal operator  (\ref{so1}) where $u$ and $v$ are arbitrary functions satisfying (\ref{CR}). And just conditions (\ref{CR}) are necessary and sufficient to keep the shape of the second derivative term in (\ref{H2}).

We will say two equations of type (\ref{seq}), (\ref{H2}) be
equivalent provided they can be connected via transformations
(\ref{Change1})--(\ref{change}).

These equivalence relations are rather powerful. In particular,
starting with Hamiltonian (\ref{H2}) with $f=1$ and making changes
(\ref{Change1}) and (\ref{Change2}) we obtain:
\begin{gather}\la{H5}H=\tilde p_a{\tilde f}\tilde p_a+\tilde
V\end{gather} where
\begin{gather}\begin{split}&\tilde p_1=-\ri\frac{\p}{\p \tilde u},
\quad \tilde p_2=-\ri\frac{\p}{\p \tilde v},\quad \tilde V
=V\left(x_1(\tilde u,\tilde v),x_2(\tilde u,\tilde v)\right),\\&\tilde f=\tilde u_1^2+\tilde u_2^2\equiv \tilde u_1^2+\tilde v_1^2.\end{split}\la{H7}\end{gather}

Formulae (\ref{H5}) and (\ref{H7}) give the infinite set of PDM
Hamiltonians which are equivalent to Hamiltonians with constant
masses. To  verify whether a given Hamiltonian (\ref{H2}) be equivalent to the Hamiltonian with a constant mass we are not supposed to search for a possibility to represent it in form (\ref{H5}), but can use a more convenient criteria presented in the following statement.

{\bf Proposition}. Hamiltonian (\ref{H2}) is equivalent to a Hamiltonian with a constant mass iff the corresponding function $f$ solves the following equation:
\begin{gather}(\log f)_{nn}=0\la{le1}\end{gather}
or, which is the same,
\begin{gather}ff_{nn}=f_mf_m.\la{le2}\end{gather}

{\bf Proof.} The generic form of Hamiltonian (\ref{H2}) equivalent to a constant mass Hamiltonian is given by equations (\ref{H5})--(\ref{H7}). Function $\tilde f$ by construction satisfies condition (\ref{le2}) thus this condition is necessary.

Let equation  (\ref{le2}) be satisfied, then representing $f$ in the form
\begin{gather}\la{lap}f=c\exp(\omega)\end{gather} we conclude that function $\omega$ should solve the $2d$ Laplace equation. It means that $\omega=F(x_1+\ri x_2)+G(x_1-\ri x_2)$.  Moreover, since $f$ by definition is real, $G$ should be nothing but the complex conjugated function $F$. It means that $f=\exp(F)(\exp(F))^\dag$, i.e., $f$ is a squared module  of a complex analytical function $U=\exp(F)$. Any such function can be represented as $U=\tilde u_1+\ri \tilde v_1$  with some $\tilde u$ and $\tilde v$ satisfying the Caushy-Riemann condition. Thus the Hamiltonian with such $f$ is equivalent to a Hamiltonian with constant mass. $\blacksquare$

In particular, if $f$ depends only on one of variables, say, $f=F(x_1)$, the corresponding Hamiltonian is equivalent to Hamiltonian with a constant mass iff $F=a\exp(b x_1)$ with some constants $a$ and $b$. One more important case: let $f=F(x_1^2+x_2^2)$, then it is possible to  reduce it to a constant inverse mass iff  $F$ is a power function. Finally, let the inverse mass be a product of functions depending on different variables, i.e.,   $f=F(x_1)G(x_2)$. Then it can be reduced to a constant iff $F(x_1)=C_1\exp(\nu x_1^2+\mu x_1)$ and $G(x_2)=C_2\exp(-\nu x_2^2+\lambda x_2)$ with some constants $C_1, C_2, \mu, \nu$ and $\lambda$.

The presented equivalence criteria will be multiple used in the following.

\section{Classification results}

We reduce the  classification of symmetries of $2d$ quantum mechanical systems with position dependent masses to solution of the determining equations (\ref{de11})--(\ref{de14}) with functions $u$ and $v$ satisfying conditions (\ref{CR}). Using equivalence relations (\ref{Change1}) and (\ref{Change2}) we can simplify the determining equations to the case $u=0$ and $v=1$, see Appendix 2. Such system of equations is easily solvable. Its general solution gives all non-equivalent Hamiltonians (\ref{H2}) such that the corresponding equations (\ref{seq}) admit a one parametric Lie group of symmetry.

A more sophisticated problem is to classify systems admitting more extended symmetries. To do it we are supposed additionally to solve determining equations of generic form (\ref{de11})--(\ref{de14}) which, however, include the restricted mass and potential terms found in the previous step.

Here we present the completed list of PDM Hamiltonians together with admitted symmetries, while the calculation details can be found in the Appendix:
\begin{gather}\la{so3}\begin{split}&
H=p_af(r)p_a+V(r),\\&Q_1=J=x_1p_2-x_2p_1;\end{split}\end{gather}
\begin{gather}\la{so4}\begin{split}&H=p_a(r^2+1)^2p_a-4r^2,\\&
Q_1=J, \ Q_2=(x_2^2-x_1^2-1)p_1-2x_1x_2p_2+2\ri x_1,\\&
Q_3=(x_1^2-x_2^2-1)p_2-2x_1x_2p_1+2\ri x_2;\end{split}\end{gather}
\begin{gather}\la{so5}\begin{split}&H=p_a(r^2-1)^2p_a-4r^2,\\&
Q_1=J, \ Q_4=(x_2^2-x_1^2+1)p_1-2x_1x_2p_2+2\ri x_1,\\&
Q_5=(x_1^2-x_2^2+1)p_2-2x_1x_2p_1+2\ri x_2\end{split}\end{gather}
\begin{gather}\la{t11}\begin{split}&H=p_ar^{\alpha+2}F(\varphi)p_a+
{r^\alpha}\hat V(\varphi), \\& Q_6=D=\ri\alpha t{\p_t}+x_1p_1+x_2p_2;\end{split}\end{gather}
\begin{gather}\la{t14}\begin{split}& H=p_af(r)p_a+\nu \varphi +\hat V(r),\\&Q_7=J+\nu t;\end{split}\end{gather}
\begin{gather}\la{t12}\begin{split}&H=p_a x_1^{\alpha+2}p_a+\nu x_1^\alpha,\quad \alpha\neq 1;\\&Q_8=p_2,\quad Q_9=D;\end{split}\end{gather}
\begin{gather}\la{t13}\begin{split}&H=p_ax_1^3p_a+\mu x_1+\nu x_2,\\&Q_{10}=p_2+\nu t,\quad Q_{11}=\ri t{\p_t}+x_1p_1+x_2p_2\end{split}\end{gather}
where $\varphi=\arctan\left(\frac{x_2}{x_1}\right)$ and $r=\sqrt{x_1^2+x_2^2}.$ In addition,
 $f(r),\ F(\varphi)$, $\hat V(r)$ and $\hat V(\varphi)$ are arbitrary functions
of the arguments fixed in brackets.

Equations (\ref{so3})--(\ref{t13}) present Hamiltonians with different symmetries, defined up to equivalence relations (\ref{Change1}), (\ref{Change2}) and  (\ref{change}). All  presented systems are invariant w.r.t. shifts of the time variable. The generator of these transformations is $P_0=\ri\frac{\p}{\p t}$.

Let us note that for $F(\varphi)=\exp(\nu\varphi)$ and $\hat V(\varphi)=\mu\exp(\nu\varphi)$ the system (\ref{t11}) admits the additional symmetry $Q=J+2\nu tP_0$. In addition, if in equations (\ref{t14}) function $f(r)$ is proportional to $r^2$, the corresponding system admits the additional integral of motion $D_0=x_1p_1+x_2p_2-\ri$. We do not specify these systems in the list (\ref{so3})--(\ref{t13}) since they are equivalent to systems with constant masses.

Operators $Q_1, \ Q_2,...,Q_5$ given by formulae (\ref{so3})--(\ref{so5}) commute with Hamiltonians and so are integrals of motion. In addition, they satisfy the following commutation relations:
 \begin{gather}\la{cr1}[Q_1,Q_2]=\ri Q_3,\quad [Q_1,Q_3]=-\ri Q_2,\quad
 [Q_2,Q_3]=\ri Q_1\end{gather}
 and
\begin{gather*}[Q_1,Q_4]=\ri Q_5,\quad [Q_1,Q_5]=-\ri Q_4,\quad
 [Q_4,Q_5]=-\ri Q_1\end{gather*}
 which specify algebras $so(3)$ and $so(1,2)$ correspondingly.

 The other symmetries which are given by equations (\ref{t11})--(\ref{t13}) are time dependent and do not commute with Hamiltonians. However, they satisfy the following relations:
  \begin{gather}\la{t15}[Q_6,H]=2\ri H,\end{gather}\begin{gather}\la{t160}[Q_7,H]=-\nu I,\quad  [Q_7,I]=[H,I]=0,\end{gather}\begin{gather}\la{t17}[Q_8,H]=0,\quad [Q_9,H]=\ri \alpha H,\quad [Q_9,Q_8]=\ri Q_8,\end{gather}
 \begin{gather}\la{t18}\begin{split}& [Q_{10},H]=-\ri\nu I,\quad [Q_{11},H]=\ri H,\quad [Q_{10},Q_{11}]=\ri \nu Q_{10},\\&[Q_{10},I]=[Q_{11},I]=[H,I]=0\end{split}\end{gather}
 where $I$ is the unit operator.

 Commutation relations (\ref{t160})  specify the Heisenberg algebra.  In the case
 (\ref{t18}) we have the Lie algebra of the centrally extended Galilei group in $1d$ space.  Relations (\ref{t15}) and (\ref{t17}) characterize the two and three dimensional solvable Lie algebras.

\section{New symmetry for systems with constant mass \la{neww}}
All systems given by formulae (\ref{so3})--(\ref{t13})  essentially differ from the systems with constant masses since the related functions $f$ do not satisfy conditions (\ref{le2}) and so cannot be reduced to constants via transformations (\ref{Change1}), (\ref{Change2}). However, if we suppose that arbitrary functions $f$ present in (\ref{so3}) and  (\ref{t14}) satisfy this condition and function $F(\varphi)$ in (\ref{t11}) is constant, we can transform the corresponding systems to constant mass ones.

Making this action with Hamiltonians (\ref{so3}) and (\ref{t11}) we come to well known systems classified in paper \cite{Boy}.  Rather surprising, it is not the case for Hamiltonian (\ref{t14}). Setting $f=1$ we obtain the following Hamiltonian:
\begin{gather}H=p_ap_a+\nu \arctan\left(\frac{x_2}{x_1}\right)+V(r)\la{new}\end{gather}
which, in addition to the obvious invariance with respect to shifts of the time variable, admits one more symmetry whose generator is
\begin{gather}\la{QQ}Q=x_1p_2-x_2p_1-\nu t.\end{gather}

Hamiltonian (\ref{new}) is missing in the classification results presented in paper \cite{Boy}. Thus our analysis of symmetries of PDM systems helped to make a small correction to this classical paper.

For completeness let us present finite group transformation generated by operator (\ref{QQ}):
\begin{gather}\la{ft}\begin{split}&x_1\to x_1\cos \theta+x_2\sin\theta,\\&x_2\to x_2\cos \theta-x_1\sin\theta,\\&\psi\to \exp(-\ri\nu\theta t)\psi\end{split}\end{gather}
where $\theta$ is a real parameter. Invariance of equation (\ref{seq}) with Hamiltonian (\ref{new}) and more general Hamiltonian (\ref{t14}) with respect to transformations  (\ref{ft}) can be easily verified by the direct calculation.

\section{Exactly solvable system \la{new2}}
Let us consider in more detail the system whose Hamiltonian is given by equation (\ref{so4}). This system admits three integrals of motion two of which, say $Q_1$ and $Q_2$,  are independent.  This number of independent integrals of motion is maximal for $2d$ Hamiltonians and characterizes so called maximally superintegrable systems.

Let us consider the eigenvalue problem for Hamiltonian (\ref{so4})
\begin{equation}\label{eq5}
  \tilde{H}\psi =-(\p_a(1 + r^2)^2\p_a+4r^2)\psi={E}\psi,
\end{equation}
where functions $\psi$ are supposed to be square integrable and vanishing at $r=0$.

Eigenvalues $E$ can be found algebraically. Indeed, integrals of motion commute with $H$ and satisfy relations (\ref{cr1}) and so form a basis of algebra $so(3)$. Casimir operator $C$ of this algebra is equal to $Q_1^2+Q_2^2+Q_2^2$, or, using realization (\ref{so4}),
\begin{gather}\la{eq25}C =Q_1^{2}+Q_2^{2}+Q_3^{2}\equiv\frac{1}{4}({H}-4).\end{gather}
Let operators $Q_1,\ Q_2$ and $Q_3$ realize an irreducible representation of algebra $so(3)$ then, in accordance with the Shurr lemma, the Casimir operator  should be proportional to the unit one. In addition, eigenvalues $c$ of $C$ are \cite{Gelf}
\begin{gather}\la{cas}
 c = s(s + 1)
\end{gather}
where $s$ are integers or half integers.

Combining (\ref{eq25}) and (\ref{cas}) we find the admissible eigenvalues of Hamiltonian in the following form:
\begin{equation}\label{14}
{E} = n^2+3,\qquad n=2s+1=1, 2, 3, ...
\end{equation}

    Let us find eigenvalues (\ref{14}) and the corresponding eigenfunctions in a more usual way. Introducing in (\ref{eq5}) the radial and angular variables $r=\sqrt{x_1^2+x_2^2},\ \varphi=\arctan\frac{x_2}{x_1}$, and expanding the wave function via eigenvectors of operator $J=x_1p_2-x_2p_1=-\ri\frac{\partial}{\partial \varphi}$:
\begin{equation}\label{19}
 \psi(x_1,x_2) =\sum_k\phi_k(r)\exp({ik\varphi}),\quad k=0,\pm1,\pm2,...
\end{equation}
we obtain the following equation for radial functions $\phi_k(r)$:
\begin{gather}\label{16}
\left(-(r^2 + 1)^2 \left(\frac{\p^2}{\p r^2}-\frac{k^2}{r^2}\right)- \frac{(1+r^2)(1+5r^2)}{r}\frac{\p}{\p r}-4r^2\right)\phi_k
 = E\phi_k.
\end{gather}

Let us note that it is reasonable to restrict ourselves to solutions with positive values of parameter $k$. Then solutions for negative $k$ can be found by transforming  $x_1 \rightarrow x_1,  x_2 \rightarrow -x_2$, and
\begin{equation}\nonumber
  \psi(x_1, x_2) \rightarrow \psi(x_1, -x_2) ,\quad   \phi_k(r)\rightarrow \phi_{-k}(r).
\end{equation}

Square integrable solutions of equation (\ref{16}) with $k\geq0$ are:
\begin{gather}\label{17}\begin{split}&
\phi_k= C_k^nr^k(r^2+1)^{-\frac{1}{2}-\frac{1}{2}n}{_2F_1}\left(\left[\frac{1}{2}-\frac{1}{2}n, \frac{1}{2}+k-\frac{1}{2}n\right], \left[1+k\right],-r^2\right),
\end{split}
 \end{gather}
 provided $E$ is given by equation  (\ref{14}) and $k\leq n-1$. Here $_2F_1$ is the hypergeometric function and $C_k^n$ are integration constants.

\section{Discussion}

We classify all non-equivalent planar QM systems with position dependent masses, which admit non-trivial Lie symmetries. In other words we describe all such symmetries which are admissible by PDM quantum mechanical systems. In this way we extend the classical results \cite{Hag}, \cite{Nied}, \cite{And} and \cite{Boy} which were restricted to the constant mass systems. Moreover, we discover a system missing in the Boyer classification results \cite{Boy}, see Section \ref{neww}. Let us note that there are $3d$ analogues of system (\ref{new}) which also are not presented in \cite{Boy}.

The presented results can be treated as certain group-theoretical grounds of the 2d quantum mechanics with position dependent masses. The ad hoc  notion of all potentially admissible Lie groups can be used for construction  of models with a priori requested symmetries. Moreover, the PDM systems  which belong  (or are equivalent)  to the systems classified in the above, can be effectively studied and simplified using the tools of Lie theory.  In particular, for the  systems equivalent to (\ref{t11}), (\ref{t12}) and it is possible to construct the similarity solutions using their invariance with respect to the scaling transformations.

The systems which admit time independent integrals of motion commuting with Hamiltonian are given in equations (\ref{so3}), (\ref{so4}) and (\ref{so5}).
 The list of such 2d PDM systems  is rather short and includes
only three representatives one of which, namely (\ref{so3}), includes two arbitrary functions.
This result is in a good accordance with Ovsiannikov's
classification of generic linear partial differential equations of
second order with two independent variables \cite{oves}.

The systems whose Hamiltonians are specified in (\ref{so4}) and (\ref{so5}) are maximally superintegrable since any of them admits two independent integrals of motion. This property makes these systems exactly solvable. The solutions of one of them are presented in Section \ref{new2}.

Let us note that potentials present in Hamiltonians  (\ref{so4}) and (\ref{so5}) do not include coupling constants. It means that the corresponding systems can be treated as purely  kinetic ones. Such interpretation can be confirmed by transformation of these Hamiltonians to the representation (\ref{H3}) where potential $\hat V$ appears to be constant and equal to 4. Since all our potentials are defined up to  constant terms, in fact we deal with Hamiltonians (\ref{H1}) with trivial potentials,  the ambiguity parameters $\alpha=\gamma=-1/2$ and masses $m=\frac1{(r^2\pm1)^2}$. In other words, in this cases it is natural to use the  Zhu-Kroemer parametrization \cite{Zhu}. All the other parametrizations including Mustafa-Mazharimousavi \cite{mustafa} one either are equivalent to the Zhu-Kroemer ordering  or lead to the presence of a potential term in the purely kinematical Hamiltonians and so are less adequate.

We remind that in the $3d$ case there exist much more (namely,
eighteen) non-equivalent systems admitting first order integrals of motion \cite{NZ}. The essential decreasing of the number of the
non-equivalent systems in the 2d case is caused by the more strong
equivalence relations (\ref{Change1}) which are defined up to an arbitrary
analytic function. For three dimensional systems such relations are
defined by transformations belonging to the 3d conformal group
$C(3)$ \cite{NZ}.

In the present paper we restrict ourselves to group classification of $2d$ systems which are  not equivalent to systems with constant masses. The classification results for systems with constant masses were presented in \cite{Boy}, but they should be  added by system discussed in Section \ref{neww}.

Let us note that the results of our analysis present in Sections 2 and 3 are valid for systems with an arbitrary number of spatial variables. In particular, they can be used to classify
Lie symmetries of $3d$ systems with position dependent masses. The list of $3d$ PDM systems with different symmetries is  much more extended then one presented in Section \ref{neww} and includes tens of representatives.  The work at this list is in progress.

\renewcommand{\theequation}{A\arabic{equation}} %
\setcounter{equation}{0}
\appendix
\section{Appendix}
\subsection{ Simplification of determining equations}

Here we prove the equivalence of systems (\ref{de2})--(\ref{de4})
and (\ref{de5})--(\ref{de8}).

The subsystem (\ref{de5}), (\ref{de6}) is nothing but the traceless
part and the trace of tensor ${\cal F}^{ab}$. Thus this subsystem is
an algebraic consequence of (\ref{de2}) and wise versa.

Let us show that equation (\ref{de3}) can be reduced to (\ref{de7}).
To this effect we differentiate (\ref{de6}) with respect to $x_a$
and  (\ref{de5}) with respect to $x_b$ with summing  up over the
repeating index $b$. As a result we obtain the following
differential consequences:
\begin{gather}\xi^i_af_i+\xi^if_{ia}-\alpha
f_a-\frac2n\left(f_a\xi^i_i+f\xi^i_{ia}\right)=0\label{a1}\end{gather}
and
\begin{gather}\frac{2-n}n\xi^b_{ab}-\xi^a_{bb}=0.\label{a2}\end{gather}

Then we multiply all terms in (\ref{de5}) by $f_b$ and make
summation over index $b$. That gives the following algebraic
consequence:
\begin{gather}\la{a3}\frac2n \xi^i_i
f_a-\xi^a_bf_b-\xi^b_af_b=0.\end{gather}

The sum of (\ref{de3}), (\ref{a1}), (\ref{a2}) and (\ref{a3}) is
nothing but equation (\ref{de7}). Then using (\ref{de7}) we can
reduce (\ref{de4}) to equation (\ref{de8}).

\subsection{Symmetries independent on $t$}

Let us prove that formulae (\ref{so3})--(\ref{so5}) present the completed  list of non-equivalent Hamiltonians admitting first order integrals of motion. For this purpose we
consider the important subclass of symmetry operators (\ref{so})
corresponding to $\xi^0=\alpha=0$ and time independent functions
$\xi^a$ and $\eta$. In this case our classification problem is
reduced to finding all nonequivalent  integrals of
motion of the following form:
\begin{gather}\la{so2}\tilde Q=up_1+vp_2-\ri u_1+\eta,\end{gather} which commute with Hamiltonian (\ref{H2}).

The related equations (\ref{de1}) turn into identities, whilst the
system (\ref{de11})--(\ref{de14}) is reduced to the following equations:
\begin{gather}\la{de15}\eta_a=0,\\\la{de16}uf_1+vf_2=2u_1f,\\\la{de17}
uV_1+vV_2+u_{11}f_1+v_{22}f_2=0.\end{gather}

In accordance with (\ref{de15}) $\eta$ should be a constant. The
corresponding term in (\ref{so2}) represents the evident constant
integral of motion. We will not consider such integrals and set
$\eta=0$.

Let Hamiltonian (\ref{H2}) commutes with  first order
differential operator (\ref{so2}). Applying  the transformation
of generic form (\ref{Change1}) we can reduce (\ref{so2}) to the
shift generator (see Appendix A3)
\begin{gather}\la{shift}\tilde Q\to\tilde Q'=-\ri\frac{\p}{\p
\tilde x_2}.\end{gather}

The corresponding transformed Hamiltonian (\ref{H5}) should commute
with (\ref{shift}). Thus
\begin{gather}\la{fv}\tilde f=\tilde f(\tilde x_1),\quad \tilde V=\tilde
V(\tilde x_1),\end{gather} i.e., both $\tilde f$ and $\tilde V$  are arbitrary functions of $\tilde x_1$.

Let Hamiltonian (\ref{H5}), (\ref{fv}) admits one more first order
integral of motion. Its generic form can be obtained from
(\ref{so2}) by changing  $u_a\to \tilde u_a$ and setting $\eta=0$:
\begin{gather}\la{so31}\hat Q=u\tilde p_1+v\tilde p_2- \ri u_1\end{gather}
where $u$ and $v$ are functions of $\tilde x_1$ and $\tilde x_2$
satisfying  relations (\ref{CR}) and  (\ref{de16}), (\ref{de17})
together with functions (\ref{fv}). The latter two relations take
the following forms:
\begin{gather}\la{de18}u\tilde f_1=2u_1\tilde f,\\\la{de171}
u\tilde V_1+u_{11}\tilde f_1=0.\end{gather}

Integrating (\ref{de18}) with respect to $x_1$ we obtain:
\begin{gather}u^2=\varphi(\tilde x_2)\tilde f(\tilde x_1)\la{de19}\end{gather}
where $\varphi(\tilde x_2)$ is a function of $\tilde x_2$. Thus $u$
should be a product of functions dependent on $\tilde x_1$ and
$\tilde x_2$:
\begin{gather}\la{de20}u=g(\tilde x_1)h(\tilde x_2).\end{gather}

In accordance with (\ref{CR}) $u$ should satisfy the equation
$u_{11}+u_{22}=0$, i.e., $g_{11}h+gh_{22}=0$. Separating variables,
we come to the following conditions for $g$ and $h$:
\begin{gather}\la{cva}\frac{g_{11}}g=-\frac{h_{22}}h=\lambda\end{gather}
where $\lambda$ is a constant.

There are three qualitatively different solutions for equations
(\ref{cva}) corresponding to the following versions of $\lambda$:
\begin{gather}\la{cvb}\lambda=0,\quad \lambda=\omega^2>0,\quad
\lambda=-\nu^2<0.\end{gather}

In the first case equations (\ref{cva}) are solved by the following
functions
\begin{gather}\la{1}g=a\tilde x_1+b, \quad h=c\tilde
x_2+d\end{gather} where $a,\ b,\ c$ and $d$ are integration
constants. In accordance with (\ref{de171}) the corresponding
potential $\tilde V$ is constant.

 Up to simultaneous shifts
and scalings of independent variables $\tilde x_1$ and $\tilde x_1$
(such transformations keep the general form of Hamiltonian
(\ref{H5})),  functions $f$ and $\tilde V$ satisfying equations
(\ref{de18}) and (\ref{de171}) can be chosen in the form
\begin{gather}\la{2}f=\tilde x_1^2, \quad \tilde V=Const.\end{gather}

Let $\lambda=\omega^2>0$, then equations (\ref{cva}) are solved by
the following functions:
\begin{gather} g=a\sin(\omega \tilde x_1)+b\cos(\omega \tilde x_1),\quad
h=c\sinh(\omega \tilde x_2)+d\cosh(\omega \tilde
x_2).\la{3}\end{gather}

Since symmetry operator (\ref{so31}) is defined up to a
multiplication constant, arbitrary constants $a$ and $b$ can be
reduced to the form $a=\cos(C), \ b=\sin(C)$ with some constant $C$. Then
\begin{gather}g=\cos(\omega \tilde x_1-C)\equiv\cos
z_1.\la{4}\end{gather} On the other hand both constants $a$ and $b$ in (\ref{3}) are essential,  and $h$ cannot be reduced to a form with one constant.

  The corresponding functions $f$ and $\tilde V$
are obtained solving equations (\ref{de18}) and (\ref{de171}):
\begin{gather}\la{Q0}f=\sin^2(z_1),\quad V=f.\end{gather}
These solutions generate Hamiltonians (\ref{H5}) which commute with the following operators:
\begin{gather}\la{Q1}\begin{split}&Q_1=\frac{\p}{\p z_2},\quad  Q_2=\sin(z_1)\cosh(z_2)\frac{\p}{\p z_1}+\cos(z_1)\sinh(z_2)\frac{\p}{\p z_2}+\cos(z_1)\cosh(z_2),\\&Q_3=\sin(z_1)\sinh(z_2)\frac{\p}{\p z_1}+\cos(z_1)\cosh(z_2)\frac{\p}{\p z_2}+\cos(z_1)\cosh(z_2)\end{split}\end{gather}
where we denote $\omega x_2=z_2.$

Analogously, considering solutions of equations (\ref{cva}) corresponding to $\lambda=-\nu^2$ we obtain the following expressions for functions $g(x_1)$ and $h(x_2)$:
\begin{gather}g=\cosh(z_1), \quad h=a\cos(z_2)+b\sin(z_2)\la{Q2}\end{gather}
and
\begin{gather}g=\sinh(z_1), \quad h=a\cos(z_2)+b\sin(z_2)\la{Q3}\end{gather}
where $z_1=\nu \tilde x_1$ and $z_2=\nu \tilde x_2.$
The corresponding solutions of equations (\ref{de18}), (\ref{de171})
are
\begin{gather} \tilde f=\cosh^2(z_1), \quad \tilde V= -f\la{Q4}\end{gather}
and
\begin{gather} \tilde f=\sinh^2(z_1), \quad \tilde V= -f\la{Q5}\end{gather}
respectively.

Notice that all Hamiltonians with the inverse masses given by equations (\ref{Q0}), (\ref{Q5}) and (\ref{2}) are equivalent. This equivalence is almost evident. Indeed, writing Hamiltonian (\ref{H5}), (\ref{2}) in circular variables $\tilde x_1=z_2\sin(z_1), \tilde x_2 =z_2\sin(z_1)$,  we immediately   reduce it to the Hamiltonian with inverse mass and potential given by equation (\ref{Q5}). On the other hand, introducing in (\ref{H5}), (\ref{2}) hyperbolical variables $\tilde x_1=z_2\sinh(z_1), \tilde x_2 =z_2\sinh(z_1)$ we obtain Hamiltonian (\ref{H5}) with constituents (\ref{Q0}).

Thus there are three non-equivalent Hamiltonians (\ref{H5}) which admit different time independent constants of motion, which correspond to the inverse masses and potentials given by equations (\ref{fv}), (\ref{Q4}) and (\ref{Q5}). Just these Hamiltonians are represented in equations (\ref{so3})--(\ref{so5}) where we introduce Cartesian  variables $x_1=\log(z_1)\cos(z_2)$ and $x_2=\log(z_1)\sin(z_2)$.

\subsection{Time dependent symmetries}
Consider now determining equations (\ref{de11})--(\ref{de14}) for the case when at least one of functions $\alpha$ and $\xi^0$ is nontrivial and functions $\eta$, $u$ and $v$ can depend on time variable $t$.

 To specify the admissible dependence on $t$ we make the following steps. Using conditions (\ref{CR}) which should be valid also for  $\dot u$ and $\dot v$, and supposing $\dot u$ and $\dot v$ to be non-trivial we obtain the following evident differential consequences of
(\ref{de11}) and (\ref{de12}):
\begin{gather}\la{t4}\dot u\hat f_2-\dot v\hat f_1-2\dot u_2=0,\\\la{t5}
\dot u\hat f_1+\dot v\hat f_2-2\dot u_1+\alpha=0 \end{gather}
where $\hat f_a=\frac{f_a}f$.

Differentiating equation (\ref{t4}) with respect to $x_2$, equation (\ref{t5}) with respect to $x_1$ and summing up the resulting expressions we obtain:
\begin{gather}\la{t16}\dot u \hat f_{nn}=0,\qquad \text{or}\qquad \dot u(\log  f)_{nn}=0.\end{gather}

It follows from (\ref{t16}) that if $u$ and $v$ are time dependent, then the  Hamiltonian admitting the corresponding symmetry should be equivalent to a constant mass one, see Proposition proved  in Section 4.1. Thus to classify essentially PDM systems it is sufficient to restrict ourselves to equations (\ref{de11})--(\ref{de14}) with $u$ and $v$ independent on $t$. In this case equations (\ref{de11}) and (\ref{de12}) are reduced to the following conditions:
\begin{gather}\la{concon}u=u(x_1,x_2),\qquad v=v(x_1,x_2),\qquad \eta=\eta(t),\end{gather}
 and the classification problem is reduced to solving equation (\ref{de13}) for $f$ and  (\ref{de14}) for $V$.

Like in previous subsection we can
reduce the terms with spatial derivatives in (\ref{so2}) to the shift generator.   Then generator $Q$ takes  the following form:
\begin{gather*}Q=\ri\xi^0\frac{\p}{\p t}+p_2+\eta.\end{gather*}
In other words we start with $u=0$ and $v=1$ which reduce equations (\ref{de13}) and (\ref{de14})  to the following forms:
\begin{gather}\la{t1} \frac{\p f}{\p x_2}=\alpha f,\quad \frac{\p V}{\p x_2}=\alpha V+\dot\eta.\end{gather}

In order to this system be consistent, $\alpha$ should be a constant and $\eta $  be a linear function of $t$ or a constant, i.e., $\eta=\nu t+\mu$, since $f$ and $V$ are time independent. Then integrating (\ref{t1}) we obtain:
\begin{gather}\la{t2}f=\exp(\alpha x_2)\tilde f(x_1), \quad V=\exp(\alpha x_2)\hat V(x_1)-\frac\nu\alpha,\quad \alpha\neq0\end{gather}
or
\begin{gather}f=\tilde f(x_1),\quad  V=\mu x_2 +\hat V(x_1), \quad \alpha=0.\la{t3}\end{gather}

Let our system admits one more symmetry of generic form (\ref{so1}). The corresponding functions $u$ and $v$ should satisfy condition (\ref{de13}) together with functions $f$ specified in (\ref{t2}) or (\ref{t3}).

 Substituting $f$ (\ref{t2})  into equation (\ref{de13}) we obtain:
\begin{gather}u\tilde f_1+\alpha v\tilde f=(\tilde \alpha+2u_1)\tilde f.\la{t51}\end{gather}

Since $u, v, \tilde f$ and $\alpha$ are time independent, new element $\tilde \alpha$ corresponding to the second symmetry should be time independent too. Moreover, considering systems admitting two symmetries, we can a priory reduce one of parameters, $\alpha$ or $\tilde \alpha$, to zero, which can be achieved by coming to linear combinations of these symmetries. We will set $\alpha=0$, i.e., restrict ourselves to  the version represented in (\ref{t3}). The corresponding  equations (\ref{t51}) and (\ref{de14}) are:
\begin{gather}\la{t9}u\tilde f_1=(\tilde \alpha+2u_1)\tilde f,\\\la{t10} u\hat V_1+\mu v+u_{11}\tilde f_1=\tilde \alpha(\mu x_2+\hat V)+\dot\eta.\end{gather}

Differentiating (\ref{t9}) with respect to $x_2$ we obtain the following condition:
\begin{gather}\la{t6} u_2\tilde f_1=2u_{12}\tilde f.
\end{gather}

Since $\tilde f$ does not depend on $x_2$, it follows from (\ref{t6}) that $u$ should have the following form:
\begin{gather}\la{t7}u=\Phi(x_1)\Psi(x_2)+\Omega(x_1)\end{gather}
with $\Phi(x_1), \Psi(x_2)$ and $\Omega(x_1)$ being functions of arguments given in the brackets.  Thus $u_2=\Phi\Psi_2$ is a product of functions of different arguments.

Since $u_2$ should satisfy the 2$d$ Laplace equation, $\Phi_{11}\Psi_2+\Phi\Psi_{222}=0,$ and so
\begin{gather}\la{t8}\frac{\Phi_{11}}\Phi=-
\frac{\Psi_{111}}{\Psi_1}=\lambda\end{gather}
where $\lambda$ is a parameter satisfying one of conditions (\ref{cvb}). Considering these conditions consequently it is not difficult to find the corresponding functions $u$ and then functions $f$ satisfying (\ref{t5}) and (\ref{t7}).  As a result we obtain Hamiltonians and the corresponding symmetries presented in equations (\ref{t11})--(\ref{t13}).

\subsection{Reduction of symmetry operator}
Let us consider the generic first order integral of motion  (\ref{so2}) and show a simple way to reduce it to the shift generator.
 The very possibility of such reduction is an element of common knowledge, see, e.g., \cite{olver}. However, we are restricted to using of special changes of variables (\ref{Change1})  which generates some specificities of this standard procedure which are presented here.

 To transform a given operator $Q$ (\ref{so2}) to the simple form  (\ref{shift}) it is necessary to change independent variables $x_1$ and $x_2$ to new variables $\tilde x_1=\varphi_1(x_1,x_2)$ and $\tilde x_2=\varphi(x_1,x_2)$, satisfying the following system of partial differential equations:
 \begin{gather}\la{aa1}Q\tilde x_1=0,\quad Q\tilde x_2=1.\end{gather}
 However, using the fact that  in our case functions $\tilde x_1$ and $\tilde x_2$ satisfy also Caushy-Riemann conditions, it  is possible to reduce (\ref{aa1}) to the following  form:
 \begin{gather}\la{aa2}\frac{\p \tilde x_1}{\p x_1}=\frac{\p \tilde x_2}{\p x_2}=-\frac{u}{u^2+v^2},\qquad \frac{\p \tilde x_1}{\p x_2}=-\frac{\p \tilde x_2}{\p x_1}=-\frac{v}{u^2+v^2}.\end{gather}

 For given functions $u$ and $v$  the system (\ref{aa2}) is easily integrable. Any its (even particular)  solution presents new variables $\tilde x_1$ and $\tilde x_2$ which which can be used to reduce the given operator (\ref{so2}) to the simple form given by equation  (\ref{shift}).

\end{document}